\documentclass[final,1p,times]{elsarticle} 
\usepackage{graphicx} 
\usepackage{amssymb} 
\usepackage{amsthm} 
\usepackage{lineno} 

\def\sqrtsNN{\mbox{$\sqrt{s_\mathrm{NN}}$}}

\newcommand{ \density} {1/{\rm S} \, {\rm d}N/{\rm dy}}
\newcommand{ \vtwoEcc} {v_2/\varepsilon}
\newcommand{ \vtwofour }{{v_2\{4\}}}
\newcommand{ \etaS }{\eta/{\rm s}}
\newcommand{ \be }{\begin{equation}}    
\newcommand{ \ee }{\end{equation}}    
\newcommand{ \bea }{\begin{eqnarray}}    
\newcommand{ \eea }{\end{eqnarray}}

\usepackage{color}

\journal{Nuclear Physics A} 
\begin{document} 

\begin{frontmatter} 


\title{Footprints of the (Nearly) Perfect Liquid}

\author{Aihong Tang}

\address{Physics Department, P.O. Box 5000, Brookhaven National Laboratory, 
Upton, NY~11973, aihong@bnl.gov}

\begin{abstract} 
In relativistic heavy-ion collisions, the system has gone through a series of 
evolution, almost at every stage of its evolution it leaves behind footprints in 
flow observable. Those footprints contain valuable information of the bulk property 
of the (nearly) perfect liquid. By examing footprints of the nearly perfect liquid, 
we address a few important issues, including the ideal hydrodynamic limit, estimation of $\etaS$, 
testing the Number of Constituent Quark scaling at low energy, in small system, at large
transverse momentum, and in forward region. Future prospect of flow study is discussed.
\end{abstract} 

\end{frontmatter} 



\section{Introduction: the (nearly) perfect liquid}
As a unique tool to study the QCD matter under extreme conditions, RHIC has 
been successful in operations since year 2000.  The wealth of data, collected and 
analyzed in many aspects, indicates that central Au+Au collisions can be well described 
by ideal Hydrodynamics~\cite{WhitePapers}.
Indications of liquid-like behavior of the matter that 
RHIC has created came in the form of large elliptic flow.  What is more 
interesting is that, this liquid has little viscosity and acts like a perfect 
one~\cite{Teaney}. Those findings lead to the announcement of the discovery of the existence 
of a perfect liquid~\cite{PerfectLiquid}. 

In following sections, we address a few important questions which are directly
related to the understanding of the property of the (nearly) perfect liquid.

\section{Is hydrodynamic limit saturated?}

The search for fluid-type signature started 30 years ago at BEVALC. Till 2000, 
when RHIC started operation, data becomes close to hydrodynamic predictions. This can be seen by
plotting $v_2$ scaled by the initial eccentricity $\varepsilon$ as a function of 
$\density$, data points reach hydro limit in most central collisions~\cite{flow130GeV}. However, 
as our understanding advances, it is realized that when calculating the hydrodynamic
limit, there is ambiguity in the choice of initial condition~\cite{CGCInit}, uncertainties in the 
Equation of State (EoS)~\cite{HydroMult}, and hadronic dissipative effects are also not negligible~\cite{HiranoSQM08}. Thus 
it is important to revisit the issue. 

\begin{figure}
\vspace{0cm}
  \begin{center}
\resizebox*{8cm}{5cm}{
\includegraphics{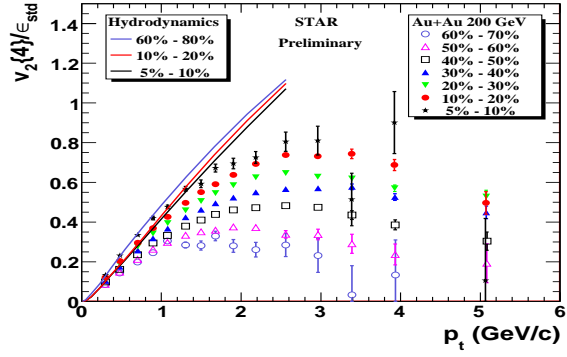}}
  \end{center}
\vspace{-0.5cm}
    \caption{ $v_2$ measured by four particle cumulant ($\vtwofour$) scaled by standard eccentricity used in hydro calculations, as a function of transverse momentum ($p_t$). Data points are for Au+Au collisions at 200 GeV. Curves are hydro dynamic predictions. Data points of this plot is from~\cite{YutingThesis}. \label{fig:v2EccHydro}}
\end{figure}

In Figure \ref{fig:v2EccHydro}, $\vtwofour$, scaled by the standard eccentricity $\varepsilon_{std}$, is plotted as a function of transvers 
momentum ($p_t$) for difference centralities. We see that $\vtwoEcc$ increases
with centrality, due to the increase of $\density$.  For $v_2$ at a 
fixed $p_t$, one finds that $v_2$ keeps increasing until it saturates at 
the bound set by hydrodynamics. This explains the saturation of $p_t$ integrated
$v_2$ in~\cite{PHENIXV2Saturation}, without the need for a softness in the EoS. \footnotemark[1].
\footnotetext[1]{
As a related remark, Heinz and Kestin pointed out at this meeting that the non-monotonic feature of $v_2$ at a fixed $p_t$
, as shown in~\cite{Kestin08}, is related to the interplay between radial flow and freeze-out. It can not be associated 
unambiguously with a phase transition in the EoS.
}

\begin{figure}
\vspace{0cm}
  \begin{center}
\resizebox*{12cm}{8cm}{
\includegraphics{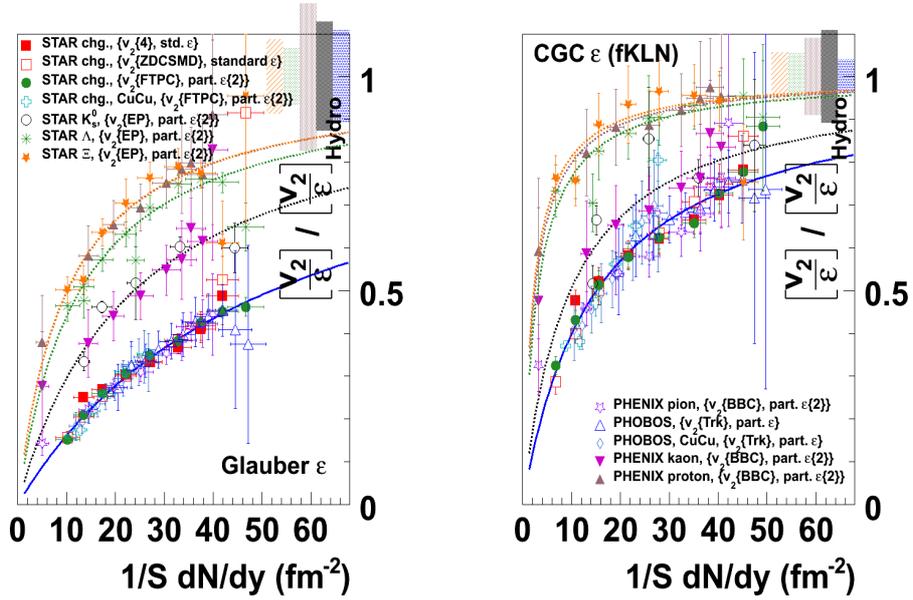}}
  \end{center}
\vspace{-0.5cm}
    \caption{ $\vtwoEcc$ scaled by the corresponding saturation value (hydro limits) obtained from the 
simultaneous fitting, for Glauber (left) and CGC (right) initial conditions. 
The hydro limit is by definition centered at unity, with error represented by 
the shaded bars. Data points are for Au+Au collisions at 200 GeV. \label{fig:pidfit}}
\end{figure}

For the $p_t$ integrated $v_2$, to quantify the possible discrepancy between
data and ideal hydrodynamics, one can either match the data using hydrodynamic
models which incorporate viscous corrections~\cite{HydroMult,Romatschke08}, or fit the data with a formula motivated
by transport models~\cite{Bhalerao:2005mm,Ollitrault_etaOverS, Raimond09}. Figure \ref{fig:pidfit} shows $\vtwoEcc$ scaled by its
corresponding limit at which they saturates (hydro limit). The hydro limit is obtained
from fitting $\vtwoEcc$ with the transported model motivated formula   $\frac{v_{2}}{\varepsilon} = \left[ \frac{v_{2}}{\varepsilon} \right]_{\rm hydro} \frac{1}{1+{\rm K/K_0}} \nonumber$, where K is the 
Knudsen number defined by the mean free path $\lambda$ scaled by the 
system size $R$, and $K_0$ is obtained from transport calculations. This formula has 
the desired feature of describing the system at two extremes; when $K$ is small, the
deviation to the saturation value (hydro limit) is proportional to K, which corresponds to 
first order correction of viscosity, and when K is large, proportional to 1/K, which 
corresponds to low density limit. A mass hierarchy is observed, the heavier the particle,
the more saturation is seen. For charged particles, we see that the system is still 30-50\%
away from ideal hydro dynamics, depending on initial conditions.

\begin{figure}
\vspace{0cm}
  \begin{center}
\resizebox*{8cm}{5cm}{
\includegraphics{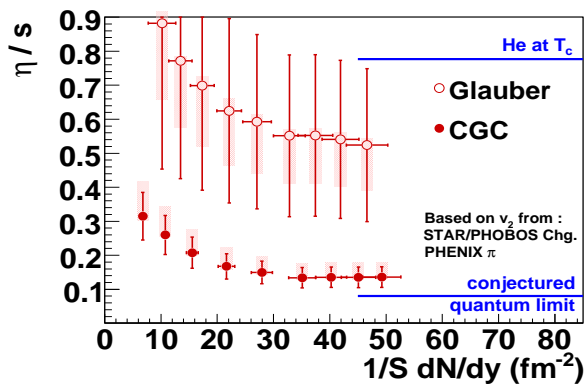}}
  \end{center}
\vspace{0cm}
    \caption{ $\etaS$ as a function of $\density$ for collisions at 200 GeV. The conjectured quantum limit, as well as $\etaS$ for He at $T_c$ is also plotted for comparison. This plot is from~\cite{Raimond09} \label{fig:viscositySTAR}}
\end{figure}

The Knudsen number obtained from this procedure can be used to calculate 
$\etaS$. Following~\cite{Teaney:2003kp}, the viscosity for a classical gas of massless
particles with isotropic differential cross sections is 
$\eta = 1.264{\rm T/\sigma}$~\cite{Kox}. 
It is arguable to apply the formula to strongly interacting dense matter,
however, in practice the viscosity recovered from this procedure agrees well
with that obtained from viscous hydro calculations~\cite{Raimond09}.
Taking the entropy density for a classical
ultra-relativistic gas as ${\rm s=4n}$, with $n$ the particle density, then  $\etaS$ can
be calculated as ${\rm \eta/s=0.316 KRT}$.
The temperature T is obtained from fitting STAR's pion ${\rm m_{T}}$ slope~\cite{STARbigSpectraPaper}.
In Figure ~\ref{fig:viscositySTAR}, $\etaS$ is plotted as a function of 
$\density$ for Glauber and CGC initial conditions. For both $\etaS$ is lower than that for He at $T_c$. 
$\etaS$ for CGC initial condition is smaller than that for Glauber initial condition, because with 
CGC initial condition, a stronger saturation is seen in the shape of 
$\vtwoEcc$ vs. $\density$, which gives a smaller K for same $\density$. This does not 
necessarily contradict to the conclusion arrived from viscous hydro 
calculations~\cite{HydroMult,Romatschke08}, in which the Equation of 
State is chosen to be the same for the two initial conditions. 

\section{What is the perfection ?}

\begin{figure}
\vspace{0cm}
  \begin{center}
\resizebox*{12cm}{8cm}{
\includegraphics{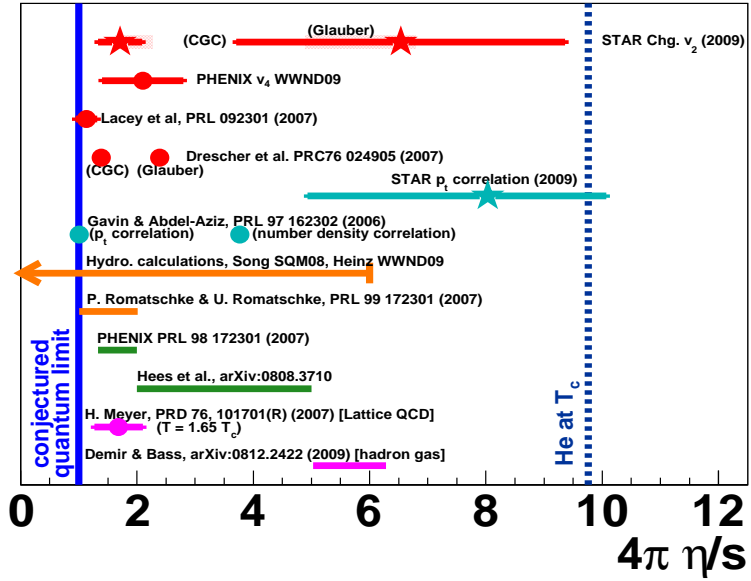}}
  \end{center}
\vspace{-0.5cm}
    \caption{ Collection of $\etaS$ calculations. Color code : Red for results based on flow measurements, cyan for results based on fluctuation measurements, orange  for results based on hydrodynamic calculations, blue for estimation based on heavy flavor measurements, and magenta for other calculations.  \label{fig:worldEtaOverS}}
\end{figure}

\begin{figure}[ht]
\vspace{0.5cm}
\begin{center}
\resizebox{
\textwidth}{!}{
\resizebox*{9cm}{6.cm}{
\includegraphics[angle=90]{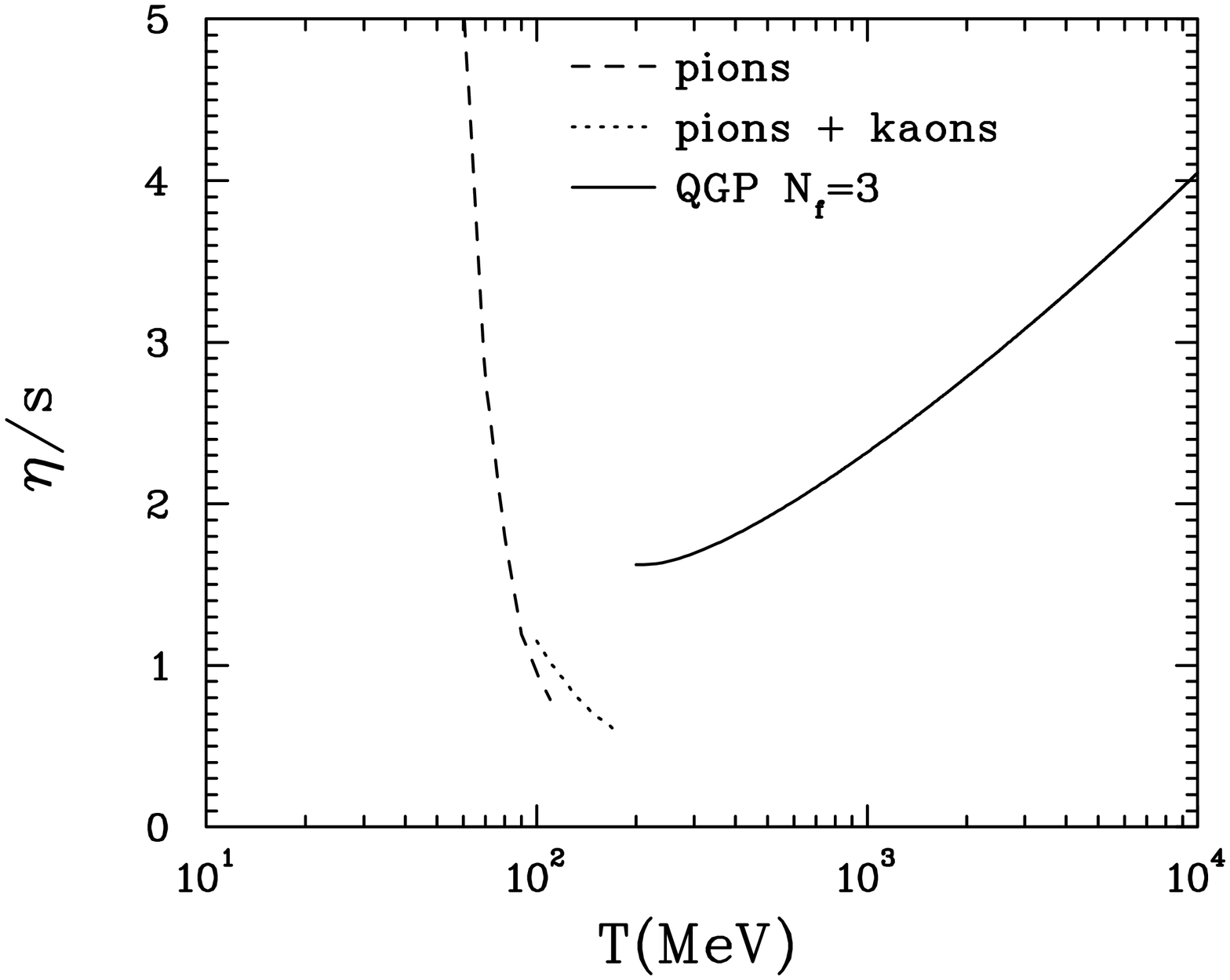}}
\resizebox*{10cm}{6.65cm}{
\includegraphics{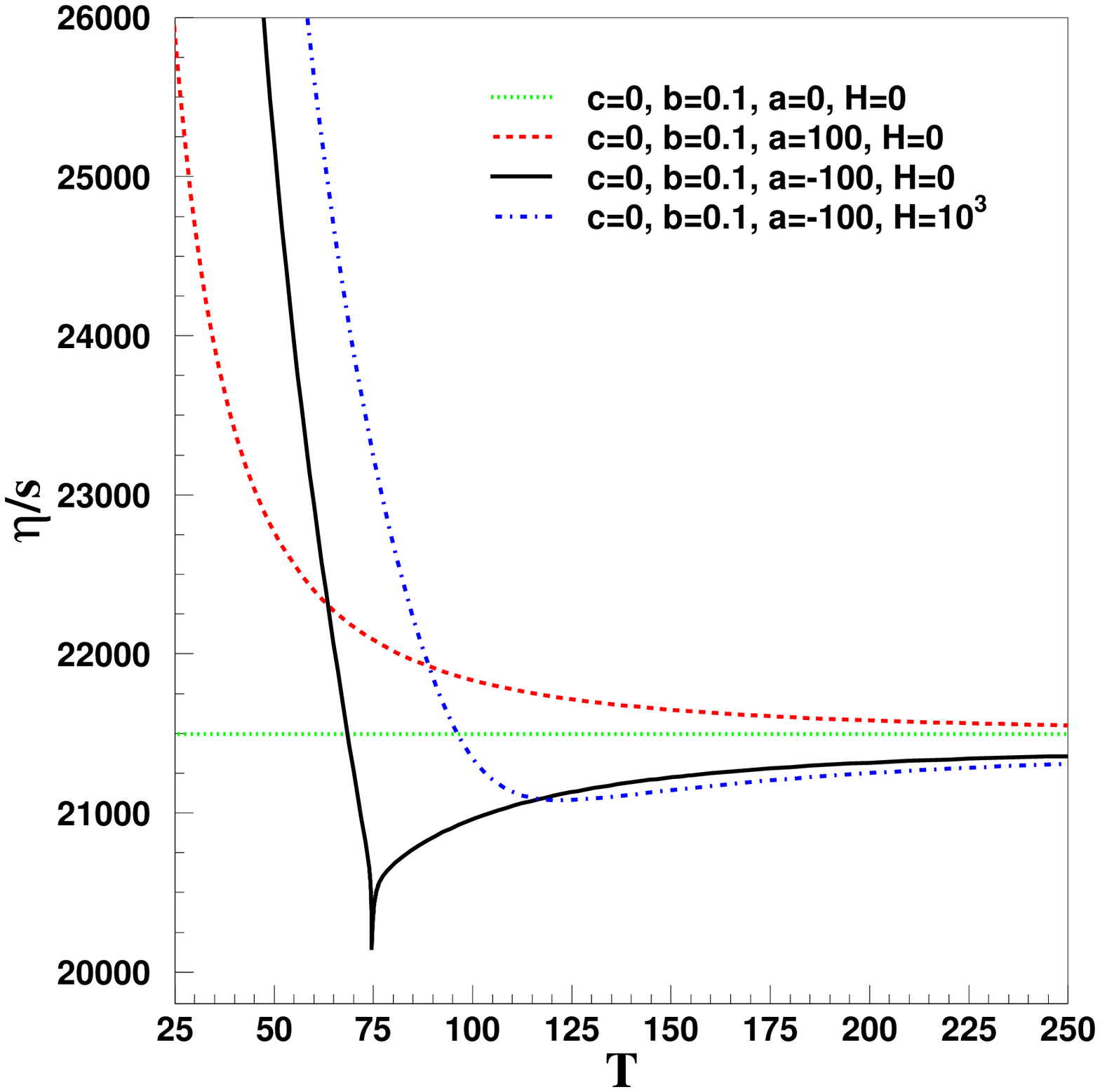}}}
\caption{ (left) $\etaS$ calculated for low energy pion gas and quark-gluon plasma~\cite{viscosCsernai}. (right) $\etaS$ calculated from weakly coupled real scalar field theories~\cite{viscosChen}. \label{fig:etaSAtTc}}
\end{center}
\vspace{-0.5cm}
\end{figure}

The $\etaS$ presented in previous section is an effective quantity  which includes viscous 
effects over different phases. It is desirable to extract $\etaS$ in QGP fluid phase,
however, all ``signatures'', spectra, flow, fluctuations, involve some time average
over the history of plasma. One has to rely on models to infer the $\etaS$ from QGP 
phase. It thus becomes important to understand the relative contribution
from phases other than the QGP fluid phase. Theoretical work~\cite{viscosCsernai, viscosChen} indicate that
above ${\rm T_c}$, the $\etaS$ increases with T, and below ${\rm T_c}$, decreases with it (Fig.~\ref{fig:etaSAtTc}). The divergence 
happens at ${\rm T_c}$ and one can infer a minimum of $\etaS$ near it. Note
that if the system spends even a short period of time in the hadronic phase it may catch 
considerable amount of viscous effect. Indeed it is calculated~\cite{DemirAndBass09} that the $\etaS$ 
for hadronic gas is 0.4-0.5, a few times of the conjectured quantum limit. Nevertheless,
if one collects $\etaS$ calculations on the market (Fig. ~\ref{fig:worldEtaOverS}), 
although the result spread in a wide range from 0 to 10 times of the conjectured quantum limit, 
which reflects the current uncertainty in the calculation of $\etaS$,
most of the calculated $\etaS$ is below that of He at ${\rm T_c}$. 

\section{What is the ultimate say on partonic collectivity ?}

\begin{figure}
\vspace{0cm}
  \begin{center}
\resizebox*{10cm}{8cm}{
\includegraphics{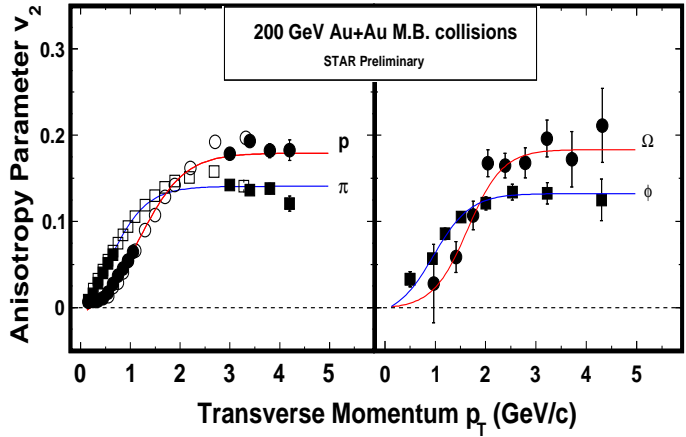}}
  \end{center}
\vspace{-0.5cm}
    \caption{ $v_2$ for light hadrons (left) and multi strange hadrons (right). Open symbols is from Phenix~\cite{PhenixPidV2} and solid symbols, STAR~\cite{Shusu09}. The curves are NCQ motivated fits~\cite{Xin04}.\label{fig:OmegaPhiV2}}
\end{figure}

$\phi$ and $\Omega$ have been suggested as an ideal probe for partonic collectivity, because 
they have small hadronic cross section thus carry information from the early, partonic stage. 
The large statistics accumulated during RHIC run VII makes it possible
to measure $v_2$ for $\phi$ and $\Omega$ with unprecedented accuracy. In the right panel of 
Figure~\ref{fig:OmegaPhiV2}, it is found that $\phi$ and $\Omega$ has sizable $v_2$ and they
still follow Number of Constituent Quark (NCQ) scaling, just like light hadrons (shown in the 
left panel for comparison). This is the definite proof that partonic collectivity has 
been reached at RHIC.

\section{What is the limit of Number of Constituent Quark scaling ?}

The flow pattern of baryons and mesons can be explained well by the 
Number of Quark scaling~\cite{NCQScaling}, which is viewed as an evidence for the existence
partonic degree of freedom. One does not expect the partonic phase to exist when 
the system is small and/or when the energy of the system is low, thus as a consequence, 
the NCQ scaling is expected to break down eventually. As a controlled reference, it would 
be interesting to find out the condition under which the NCQ scaling breaks down.

\begin{figure}[ht]
\vspace{0.5cm}
\begin{center}
\resizebox{
\textwidth}{!}{
\resizebox*{10cm}{6.1cm}{
\includegraphics{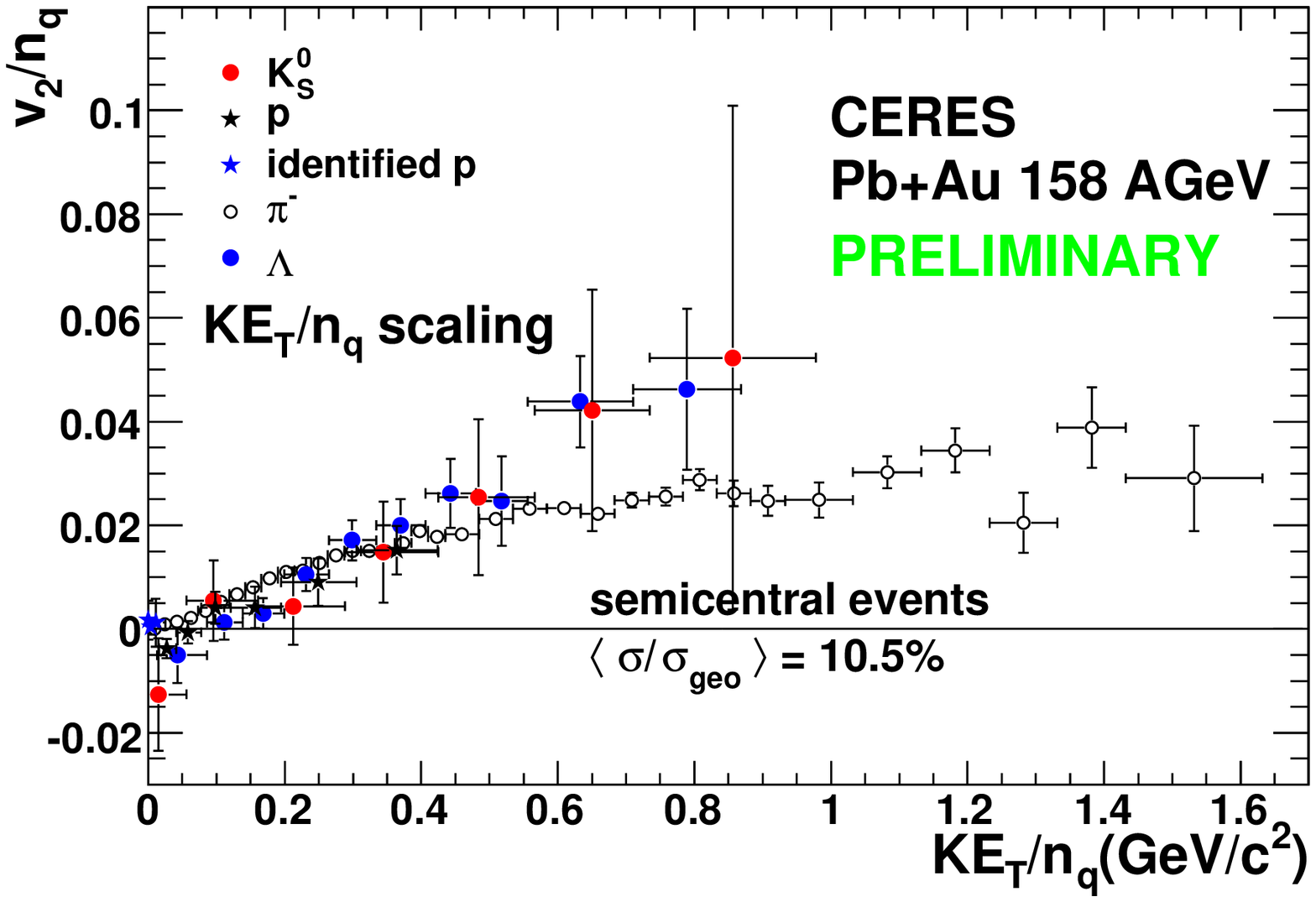}}
\resizebox*{10cm}{6.cm}{
\includegraphics{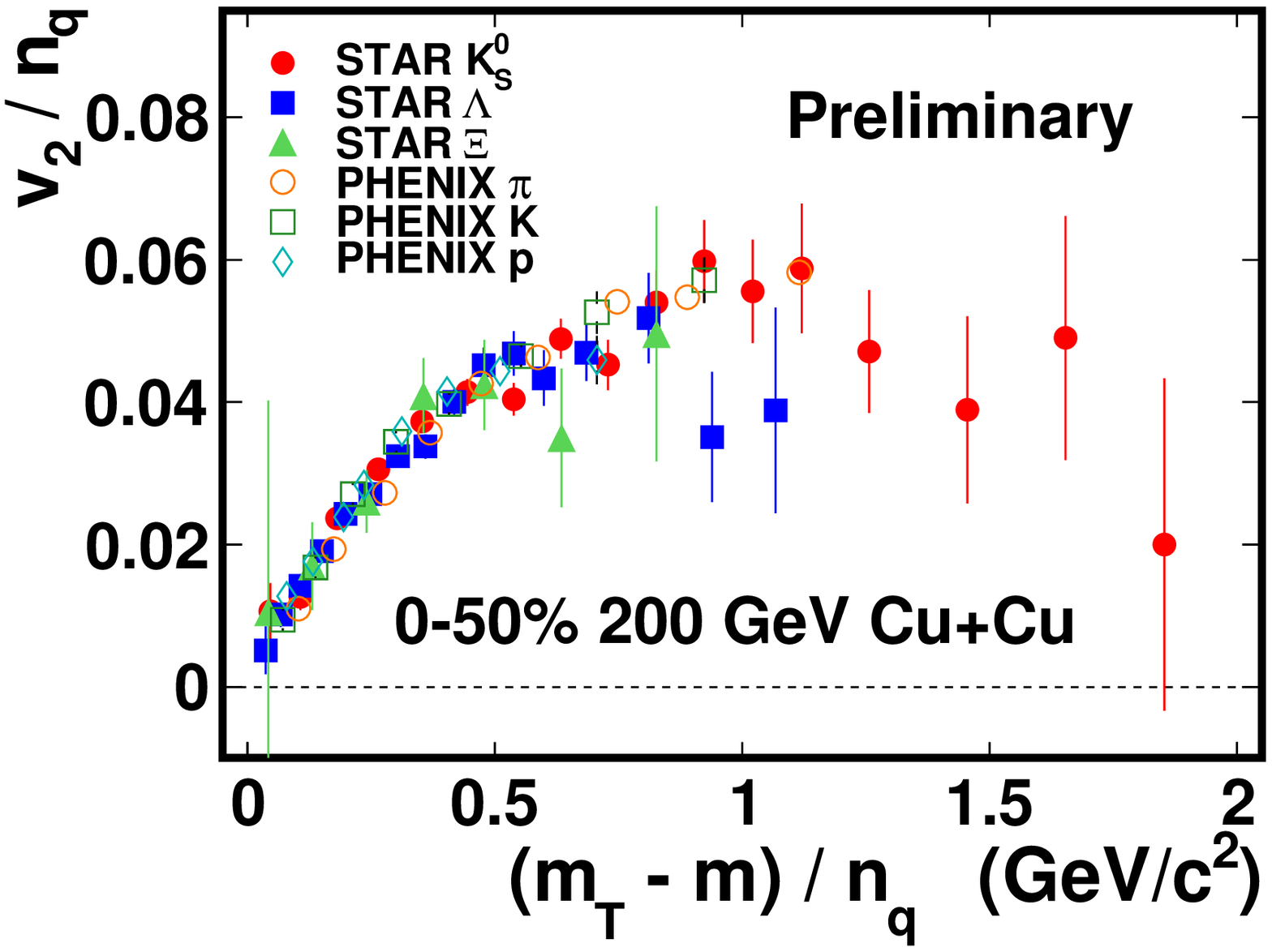}}}
\caption{ $v_2$ scaled by number of quarks ($n_q$) as a function of $kE_T/n_q$, for Pb+Au collisions at 158 AGeV~\cite{Milosevic09} (left) and CuCu collisions at 200 GeV~\cite{Shusu09} (right). \label{fig:NCQLowESmallSys}}
\end{center}
\end{figure}

In the left panel of Figure~\ref{fig:NCQLowESmallSys}, $v_2$ of identified particles
are scaled by their constituent quarks and plotted as a function of $kE_T/n_{q}$,
for Pb+Au collisions at 158 AGeV. We see that within errors, the result is still consistent with NCQ scaling up to 
$kE_T/n_{q}=0.8 \mathrm{GeV/c^{2}}$. In the right panel, a similar plot is shown
for Cu+Cu collisions at 200 GeV, $v_2/n_q$ for identified particles fall in 
a common trend well. No sign shows that NCQ breaks at these two conditions.
NCQ is also studied at forward region in~\cite{Sanders09}, within error there 
is no sign of breakdown.

Besides the partonic degree of freedom, NCQ scaling also implies that particles at 
intermediate $p_t$ are produced by quark coalescence. At large $p_t$ as hard
process begins to kick in and particles are no longer produced by quark coalescence,
thus the NCQ scaling is expected to break down. It is important to locate the
$p_t$ range where the NCQ scaling breaks down, as it tells us the region of transition of particle production mechanism. 
It is as well important to examine the pattern with which the NCQ scaling breaks down
for various hadrons. Such pattern will not only shed a light on the dynamics of jet 
fragmentation, it will also, being served as a counter example of NCQ scaling, 
deepen our understanding of quark coalescence.
\begin{figure}[ht]
\vspace{0.5cm}
\begin{center}
\resizebox{
\textwidth}{!}{
\resizebox*{10cm}{6.07cm}{
\includegraphics{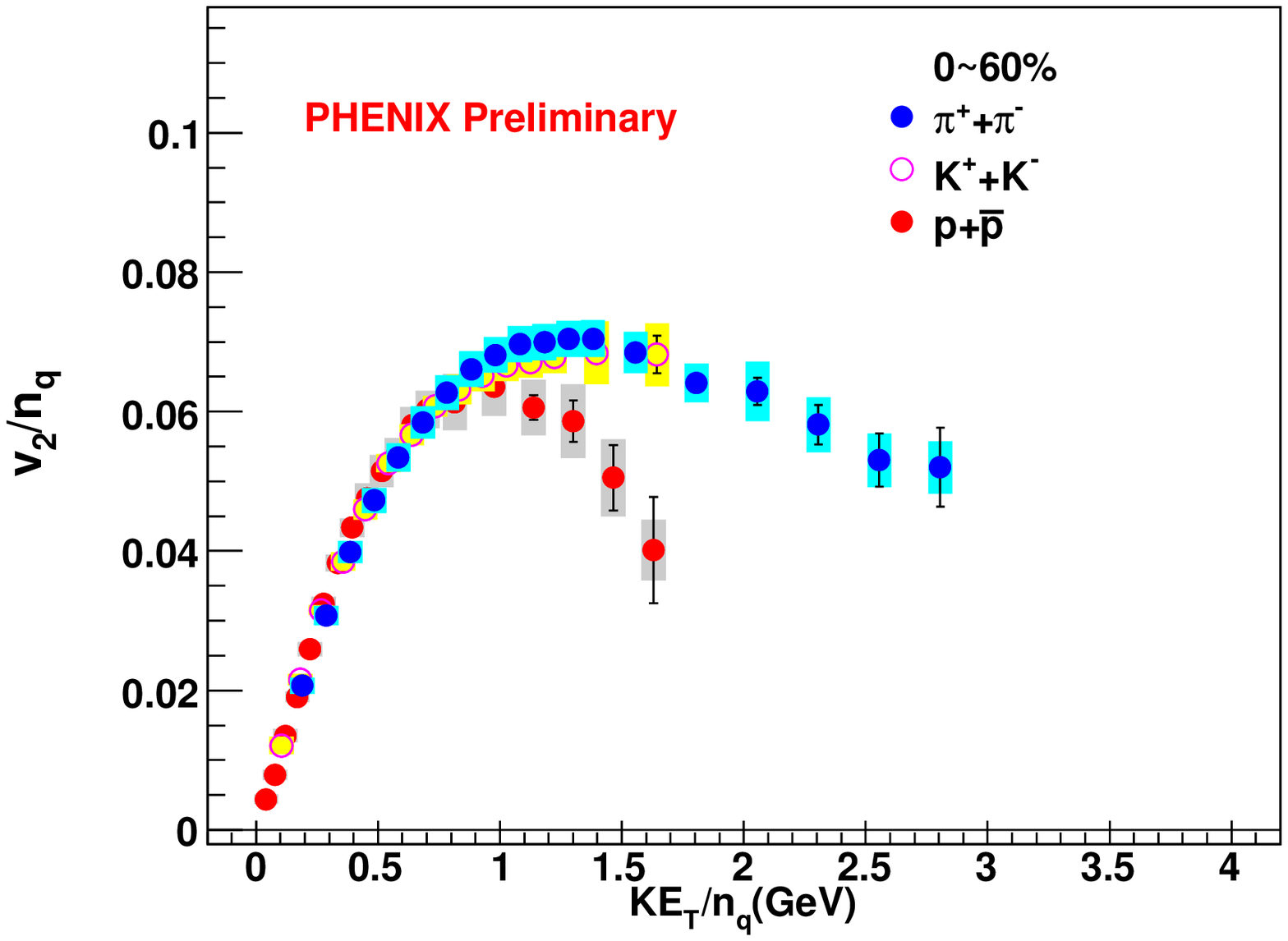}}
\resizebox*{10cm}{6.cm}{
\includegraphics{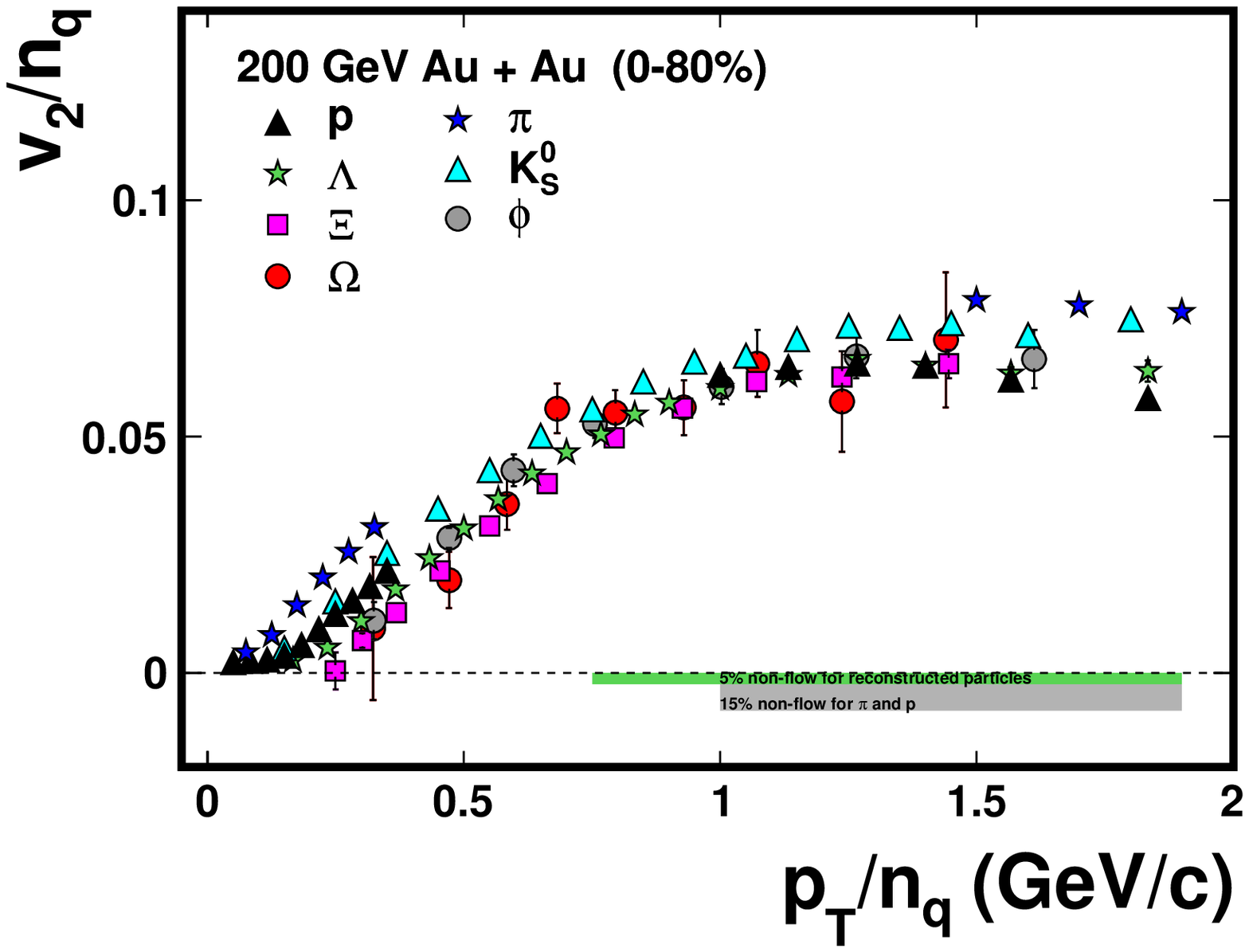}}}
\caption{ $v_2$ scaled by number of quarks ($n_q$) as a function of $kE_T/n_q$~\cite{Shengli08}(left) and $p_t/n_q$~\cite{Shusu09} (right), for Au+Au collisions at 200 GeV\label{fig:largePt_nq}}
\end{center}
\vspace{-0.5cm}
\end{figure}
Figure~\ref{fig:largePt_nq} shows $v_2/n_q$ for identified particles from
both PHENIX (left) and STAR (right). It is obvious that  $v_2/n_q$ of 
protons and pions begin to diverge starting around ${\rm kE_T/n_q=1 GeV}$, possible 
difference between lambda and pions is also observed.

\section{Outlooks}

\begin{figure}[ht]
\vspace{0.5cm}
\begin{center}
\resizebox{
\textwidth}{!}{
\resizebox*{8cm}{4.83cm}{
\includegraphics{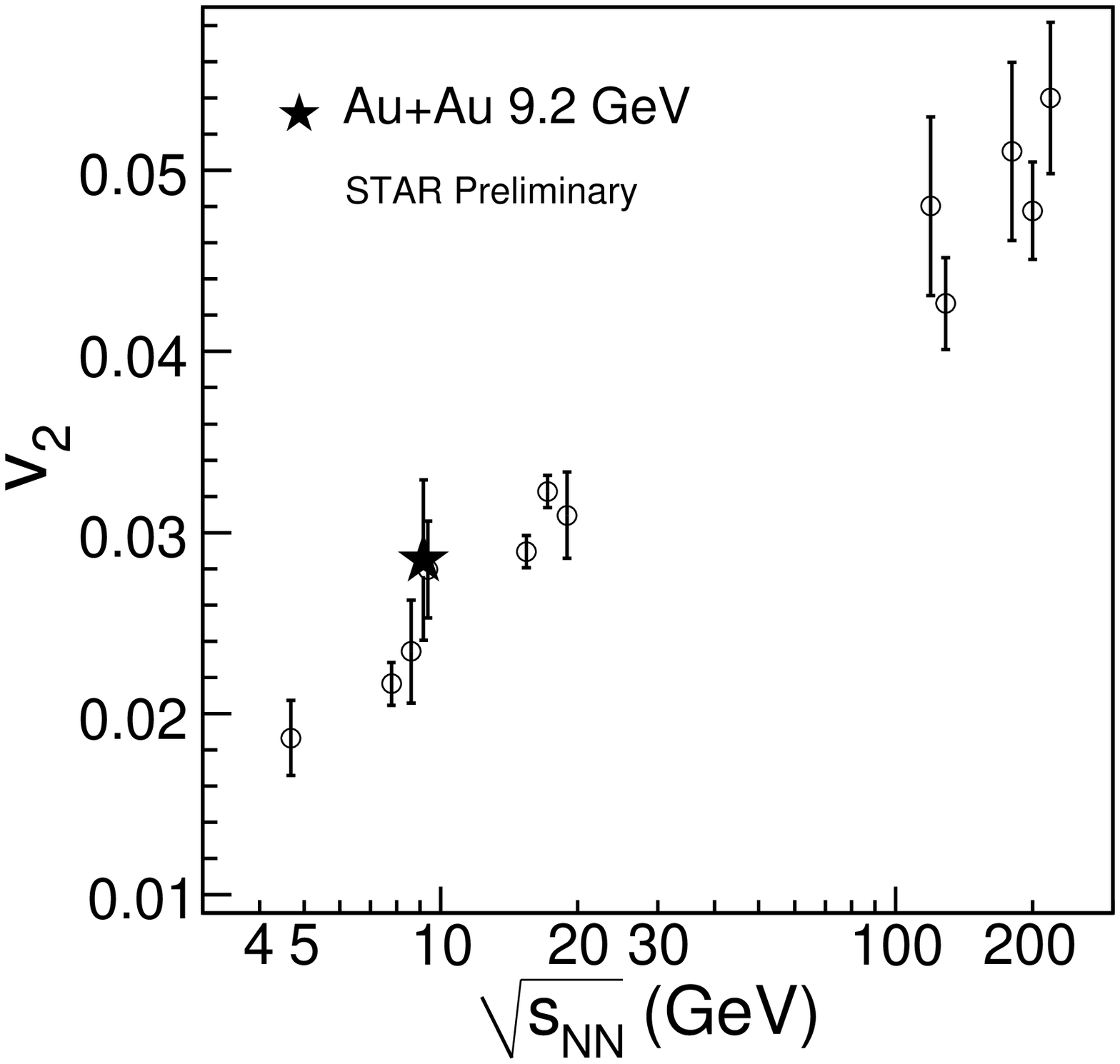}}
\resizebox*{8cm}{5.1cm}{
\includegraphics{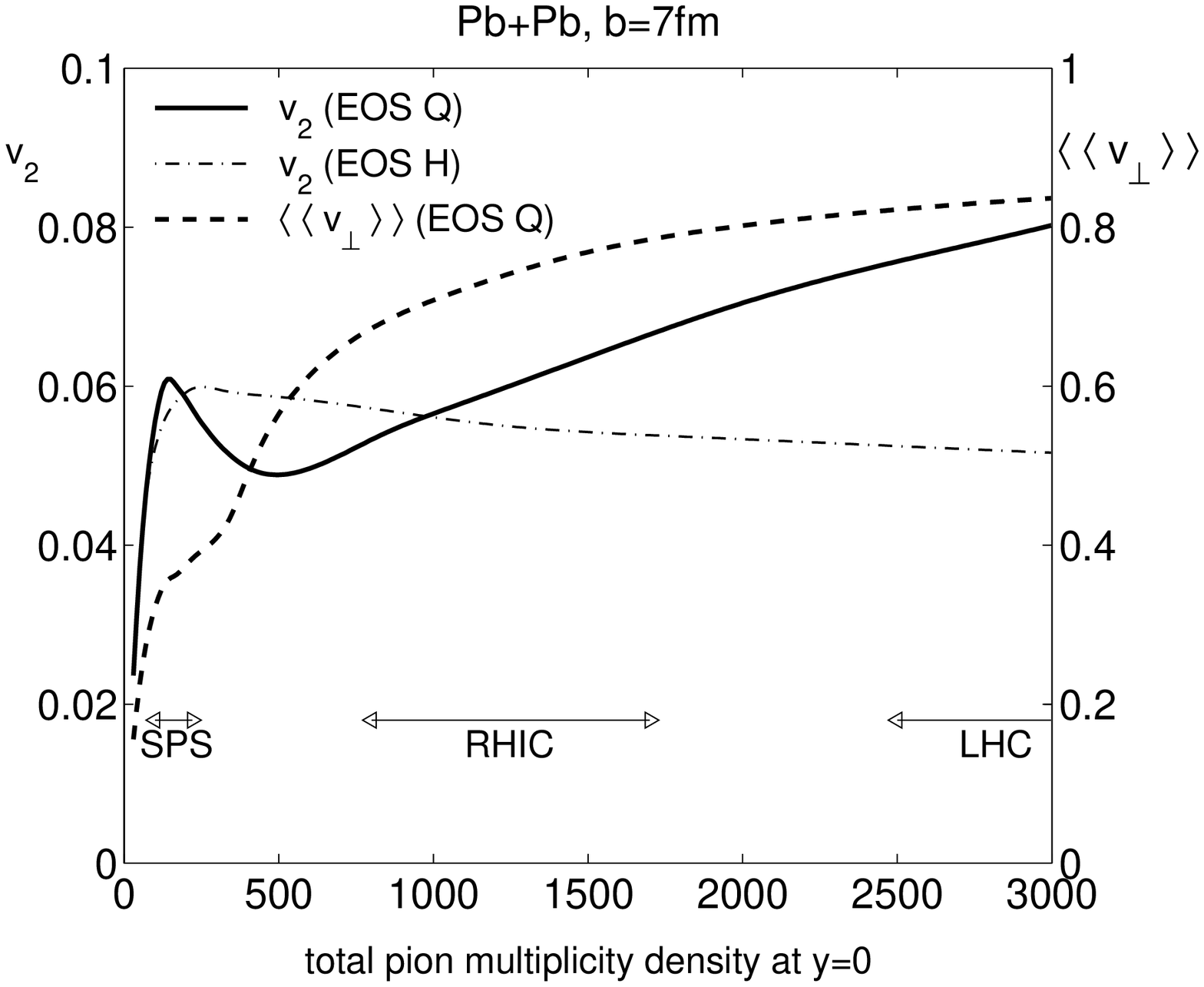}}}
\caption{Left: $v_2$ excitation function, the plot is from~\cite{Lokesh09}. Right: Hydro calculation of integrated $v_2$ as a function of pion multiplicity, for a given impact parameter. The plot is from~\cite{Kolb2000} \label{fig:BESV2}}
\end{center}
\vspace{-0.5cm}
\end{figure}

From the (3+1)-dimensional hydrodynamics with exact longitudinal boost-invariance~\cite{Kolb2000}, 
it is demonstrated that a phase transition from hadronic matter to QGP leads to
non-monotonic behavior in both beam energy and impact parameter dependence. This
is shown in right panel of Figure~\ref{fig:BESV2}, in which for a combined 
Equation of State (EOS Q) that includes the phase transition between a resonance
gas (EOS H) and for the QGP phase (EOS I), a dip is observed at the range of 
SPS energy. As $v_2$ begins to rise again for high density at LHC energies, 
the dip only covers the energy range between SPS and RHIC. The significance
and the range it happens has to be determined by experiments.
Left panel of 
Figure~\ref{fig:BESV2} shows the excitation function of $v_2$. We see a general
trend of $v_2$ increase with energy (thus density), with current error
we cannot identify any possible abnormal
around \sqrtsNN =10 Gev, the same energy at which the $k^{+}/\pi^{+}$ ``horn''~\cite{horn} is located.
It is desirable to return to this energy range with accurate measurement 
to identify possible abnormalities.  

Looking into RHIC II program and future heavy ion program at LHC, as part of a comprehensive 
test for the deviation from hydro limit, it is important to
i) increase the system size while keep the energy the same (U+U collisions),
and ii) increase the energy while keep the system size the same. The former will
allow us to test if $v_2$ saturates or not by increasing the particle density in the
transverse plane, and the latter will allow us to check if the viscosity increases
as anticipated once we passed the transition temperature.

\section{Summary}

The deviation from ideal hydrodynamic limit is discussed. From the transport model based approach, it is found that the system is 30-50\% away from the value at which it is supposed to saturate. The perfection of the nearly perfect liquid is reviewed. The current estimations of $\etaS$ spread in a wide range from 0 to 10 times of the conjectured quantum limit, however, most of the calculated $\etaS$ is below that of He at ${\rm T_{c}}$. It is concluded that the partonic collectivity has been reached at RHIC. The number of quark scaling is found to hold roughly at low energies, in small systems and in forward regions, however, signs of break down of NCQ scaling are observed at large $p_t$. To the aspect of understanding the bulk property of the nearly perfect liquid, the importance of future heavy ion programs at RHIC and LHC is emphasized.

\section*{Acknowledgments} 

The author thank following people for providing inputs/plots for this paper: S.~Esumi, P.~Fachini, U.~Heinz, T.~Hirano, S.~Huang, L.~Kumar, R.~Lacey, N.~Li, H.~Masui, J.~Milosevic, J-Y.~Ollitrault, A.~Poskanzer, S.~Sanders, M.~Sharma, S.~Shi, R.~Snellings, H.~Song, A.~Taranenko, G.~Usai, S.~Voloshin, N.~Xu.


\begin{thebibliography}{00} 


\bibitem{WhitePapers}   
BRAHMS, PHENIX, PHOBOS, and STAR Collaboration, 
{\it Nuclear Physics A} {\bf 757} (2005) Issues 1-2.

\bibitem{Teaney} 
D.~Teaney, 
{\it Phys. Rev. C} {\bf 68} (2003) 034913.

\bibitem{PerfectLiquid}
BRAHMS, PHENIX, PHOBOS, and STAR Collaboration,
{\it Hunting the Quark Gluon Plasma: Results from the First 3 Years at RHIC} (Upton, NY: Brookhaven National Laboratory report No. BNL-73847-2005).

\bibitem{flow130GeV}
C.~Adler et al., [STAR collaboration],
{\it Phys. Rev. C} {\bf 66} (2002) 034904

\bibitem{CGCInit}
A.~Adil et al., {\it Phys. Rev. C} {\bf 74} (2006) 044905 \\
T.~Hirano et al., {\it Phys. Rev. B} {\bf 636} (2006) 299.

\bibitem{HydroMult}
H.~Song and U.~Heinz, {\it Phys. Rev. C} {\bf 78} (2008) 024902.

\bibitem{HiranoSQM08}
T.~Hirano, (2008) arXiv:0812.4651.

\bibitem{PHENIXV2Saturation}
S.~Adler et al., [PHENIX collaboration], {\it Phys. Rev. Lett.} {\bf 94} (2005) 232302.

\bibitem{YutingThesis}
Y.Bai, Ph.D. Thesis. University Utrecht, the Netherlands, (2007).

\bibitem{Kestin08}
G.~Kestin and U.~Heinz, {\it Eur. Phys. J. C} {\bf 61} (2009) 545-552.

\bibitem{Romatschke08}
  M.~Luzum and P.~Romatschke,
  {\it Phys. Rev. C} {\bf 78} (2008) 034915.


\bibitem{Bhalerao:2005mm}
  R.~S.~Bhalerao, J.~P.~Blaizot, N.~Borghini and J.~Y.~Ollitrault,
  {\it Phys. Lett. B} {\bf 627} (2005) 49.
  

\bibitem{Ollitrault_etaOverS}
  H.~J.~Drescher, A.~Dumitru, C.~Gombeaud and J.~Y.~Ollitrault,
  {\it Phys. Rev. C} {\bf 76} (2007) 024905.

\bibitem{Raimond09}
  R.~Snellings, H.~Masui, J-Y.~Ollitrault and A.~Tang, (2009) this Quark Matter proceeding.

\bibitem{Teaney:2003kp}
  D.~Teaney,
  {\it Phys. Rev. C} {\bf 68} (2003) 034913.

\bibitem{Kox}
  A.~J.~Kox, S.~R~ de Groot, and W.~A.~van~Leeuwen,
  {\it Phys. A} {\bf 84} (1976) 155.

\bibitem{STARbigSpectraPaper}
  B.~Abelev et al., [STAR collaboration], {\it Phys. Rev. C} {\bf 79} (2009) 034909.

\bibitem{viscosCsernai}
  L.~Csernai, J.~Kapusta and L.~McLerran, {\it Phys. Rev. Lett.} {\bf 97} (2006) 152303.

\bibitem{viscosChen}
  J-W.~Chen, M~Huang,Y-H.~Li, E.~Nakano and D-L.~Yang, {\it Phys. Lett. B} (2008) 67018.

\bibitem{DemirAndBass09}
  N.~Demir and S.~Bass, {\it Phys. Rev. Lett.} {\bf 102} (2009) 172302.

\bibitem{PhenixPidV2}
  M.~Issah and A.~Taranenko for the PHENIX collaboration, (2006) arXiv:nucl-ex/0604011.
 
\bibitem{Shusu09}
  S.~Shi for the STAR collaboration, (2009) arXiv:0907.2265.

\bibitem{Xin04}
  X.Dong et al., {\it Phys. Lett. B} {\bf 597} (2004) 328.

\bibitem{NCQScaling}
  J.~Adams et al., [STAR collaboration], {\it Phys. Rev. Lett.} {\bf 95} (2005) 122301.

\bibitem{Milosevic09}
  J.~Milosevic for CERES collaboration, {\it J. Phys. G} {\bf 32} (2006) S97.

\bibitem{Sanders09}
  S.~Sanders for the BRAHMS collaboration, (2009) this Quark Matter proceeding.

\bibitem{Shengli08}
  S.~Huang for the PHENIX collaboration,{\it J. Phys. G} {\bf 35} (2008) 104105.

\bibitem{Kolb2000}
  P.~Kolb, J~Sollfrank and U.~Heinz, {\it Phys. Rev. C} {\bf 62} (2000) 054909.

\bibitem{horn}
  S.~Afanasiev et al., [NA49 collaboration], {\it Phys. Rev. C} {\bf 66} (2002) 054902.

\bibitem{Lokesh09}
  L.~Kumar for the STAR collaboration, arXiv:0907.1943

\end{thebibliography}
\end{document}